\documentclass[reprint,bibnotes,amsmath,amssymb,aps,prb,showpacs,floatfix,superscriptaddress]{revtex4-1}
\usepackage{graphicx}
\usepackage{color}
\usepackage{amsmath}
\usepackage{amssymb}
\usepackage[breaklinks=true,colorlinks,citecolor=blue,linkcolor=blue,urlcolor=blue]{hyperref}

\renewcommand{\BibitemShut}[1]{}
\usepackage{epsfig,mathrsfs,color,latexsym,subfigure,marginnote}

\begin{document}

\title{Electric-field tunable Dirac semimetal state in phosphorene thin films}

\author{Barun Ghosh}
\affiliation{Department of Physics, Indian Institute of Technology Kanpur, Kanpur 208016, India}

\author{Bahadur Singh}
\email{bahadursingh24@gmail.com}
\affiliation{Centre for Advanced 2D Materials and Graphene Research Centre, National University of Singapore, Singapore 117546}
\affiliation{Department of Physics, National University of Singapore, Singapore 117542}

\author {Rajendra Prasad}
\affiliation{Department of Physics, Indian Institute of Technology Kanpur, Kanpur 208016, India}

\author{Amit Agarwal}
\email{amitag@iitk.ac.in}
\affiliation{Department of Physics, Indian Institute of Technology Kanpur, Kanpur 208016, India}
\date{\today}

\begin{abstract}
We study the electric-field tunable electronic properties of phosphorene thin films, using the framework of density functional theory. We show that phosphorene thin films offer a versatile material platform to study two dimensional Dirac fermions on application of a transverse electric field. Increasing the strength of the transverse electric field beyond a certain critical value in phosphorene thin films leads to the formation of two symmetry protected gapless Dirac fermions states with anisotropic energy dispersion. The spin-orbit coupling splits each of these Dirac state into two spin-polarized Dirac cones which are also protected by non-symmorphic crystal symmetries. Our study shows that the position as well as the carrier velocity of the spin polarized Dirac cone states can be controlled by the strength of the external electric field. 

\end{abstract}
\maketitle
\section {Introduction}

Following the successful exfoliation of graphene in 2004 \cite{Novoselov666,Geim2007}, there has been a huge surge of interest in two-dimensional (2D) materials, and a large variety of 2D materials have been proposed theoretically and also synthesized experimentally, both in the compound as well as in the elemental form. Few examples include, but are not limited to, boron nitride (h-BN), transition metal dichalcogenide MX$_2$ (where M = Mo or W and X = S, Se or Te), silicene, and  germanene \cite{Geim2007,2D_review,Dubertret:2015aa,SiGe,silicene,Si2,SiGe2,stanene,TMD1,TMD2,Ghosh_2015,Rastogi:2014aa,Nahas:2016,Piyush:2016,PhysRevB.94.035423}.
Phosphorene - the few layered structure of black phosphorous, is another interesting 2D material, which has attracted attention recently following the successful fabrication of the field effect transistor using thin multilayered phosphorene \cite{P1,P0}. It is a direct band gap semiconductor, and has high carrier mobility ($\sim 10^3 $cm$^2$/Vs) along with a high on/off current ratio ($\sim10^5$), in addition to very interesting anisotropic thermal, optical, and electronic transport properties \cite{P2,P3,P4,PhysRevB.92.165406,PhysRevB.93.245433,PhysRevB.93.241404,PhysRevB.92.165406}. The energy gap as well as the carrier properties of phosphorene strongly depend on the external perturbations such as strain, electric field, pressure and dopants \cite{PhysRevLett.112.176801,Zhang2015,doi:10.1021/jp506881v,PhysRevLett.114.046801,Nahas:2016}. Due to the existence of these novel properties, phosphorene offers a promising potential in electronics, optoelectronics, and spintronic applications \cite{doi:10.1021/acs.jpclett.5b01686,doi:10.1021/acs.jpclett.5b01094}. 

Meanwhile, the study of Dirac fermions and exploring their various properties in 2D and 3D materials,  is another exciting field of research both from the perspective of the 
condensed matter physics as well as the materials science community \cite{RMP1_mzh,RMP2_qi,RMP3_AB,weyl_wan2011,weyl_singh2012,kane3d,kane2d}. The Dirac fermions state has been extensively studied in graphene and on the surface of topological insulators in 2D and in topological semimetals such as Dirac/Weyl semimetals in 3D. Notably, the Dirac fermions state in topological insulators and topological semimetals arises because of the non-trivial topology and is protected either by the time reversal symmetry or other crystal symmetries. 
However in the case of graphene, while the Dirac Fermion state is protected by crystal symmetry in the absence of spin-orbit coupling (SOC), a small but finite band gap opens up at the Dirac points in the presence of SOC \cite{qshe_graphene}. More recently, the symmetry protected 2D Dirac semimetal states has been proposed in materials with the non-symmorphic crystal symmetries \cite{kane2d}. This was followed by the theoretical prediction of 2D Dirac fermion state in group-Va elements with phosphorene lattice structure which has many non-symmorphic symmetries and among others \cite{Lu2016}. Motivated by these studies, in this article,     we explore the Dirac fermion states in Phosphorene. 
 
In a recent experimental study followed by first principle calculations, it has been shown that thin films of phosphorene with potassium doping can support 2D Dirac semimetal state on the surface of the thin films \cite{Kim723,Kdoped1}. The theoretical study has in fact shown that although spin-orbit coupling lifts the spin-degeneracy of states, it does not open a band gap at the Dirac points on account of a unique form of the SOC Hamiltonian, leading to the Dirac semimetal state \cite{Kdoped1}. Other studies have shown that the band gap in the phosphorene thin films can be modulated by an external electric field applied perpendicular to the thin films. The electric field applied beyond a critical value inverts the normal band order with the formation of two Dirac cones in the absence of SOC. However, they proposed that the SOC-induced a band gap of $\sim5$ meV at the Dirac points and thus phosphorene thin films realize a quantum spin-Hall insulator state with $Z_2=1$ \cite{nanolettntot}. Later, by analyzing the behavior of the  pseudo-relativistic Dirac cones, Dolui {\it et al.} claimed that the SOC opens up a bandgap at the Dirac points at some intermediate electric field strength \cite{Dolui2015}. The SOC induced band gap decreases with an increase in electric field and closes again to realize a Dirac semimetal state in the presence of SOC. 

The preceding discussion made it clear that thin films of phosphorene realizes an electric field (gating) induced topological phase transition into a Dirac semimetal state. However, a detailed study of the emergence of the Dirac semimetal state and its symmetry protection is still lacking. Accordingly, in this paper, we have systematically investigated the electronic properties of ultra-thin $N-$layer (NL) phosphorene within the density functional theory framework. We show that the application of a transverse electric field to NL phosphorene films causes a monotonic decrease in the  band gap, eventually leading to the formation of two Dirac cones on the high symmetry line beyond a critical field strength. Interestingly, we find that the position, as well as the associated carrier velocity of the Dirac cone states, can be uniquely controlled by an external electric field. A symmetry analysis shows that these Dirac cones are protected by non-symmorphic symmetries of the crystal lattice. The Inclusion of the SOC in presence of the transverse electric field,  splits each Dirac cone into two spin polarized Dirac cones, which are located at generic $k$-points of the 2D Brillouin zone. These Dirac cones in phosphorene show a unique spin-texture where spins are aligned almost along the {\it x}-directions. We find that the unusual form of the effective SOC Hamiltonian preserves the non-symmorphic symmetry and thus NL phosphorene realizes a symmetry protected Dirac semimetal state under an external electric field.

The article is organized  as follows: Section~\ref{computations} gives the relevant computational details and discusses the structural properties of NL phosphorene.  In Sec. \ref{QCE}, we discuss the effect of quantum confinement and thickness dependent band structure of NL phosphorene thin films. Section \ref{EField} focuses on the emergence of symmetry protected Dirac semimetal state in phosphorene under an external electric field. Finally, we summarize our findings in Sec. \ref{concl}.

\section{structural properties and computational details}
\label{computations}

Electronic structures were calculated within the framework of the density functional theory (DFT) \cite{PhysRev.140.A1133}, using the VASP (Vienna Ab Initio Simulation Package) suit of codes. We used projector augmented wave (PAW) potentials with an energy cut-off of 500 eV for the plane wave basis set and a tolerance of $10^{-6}$ eV for electronic energy minimization \cite{paw,paw1}. 
The exchange-correlation effects were included using a generalized gradient approximation (GGA) with the van der Waal's correction (PBE + vdw-DF2)\cite{vasp,vdw1,vdw2,pbe}. We used the conventional bulk unit cell of black phosphorous with relaxed structural parameters, $a=3.39$ \r{A},  $b=4.73$ \r{A} and $c=11.33$ \r{A}, to construct NL phosphorene thin films. A vacuum of $10$~\r{A} was used on both sides of the slab to avoid interaction between spurious replica images. The Brillouin zone sampling was done using $\Gamma$-centered 12$\times$12$\times$8 and 12$\times$10$\times$1 $k$-meshes for the bulk and thin films of phosphorene, respectively. All the atomic positions were relaxed until the residual forces on each atom were less than  $1.0\times10^{-3}$eV/\r{A}. To simulate the effect of the electric field, a sawtooth-type periodic potential was applied perpendicular to the thin films. The spin-orbit coupling (SOC) was added self-consistently to include relativistic effects \cite{PhysRevB.79.224418}.

\begin{figure}
\includegraphics[width=0.50\textwidth]{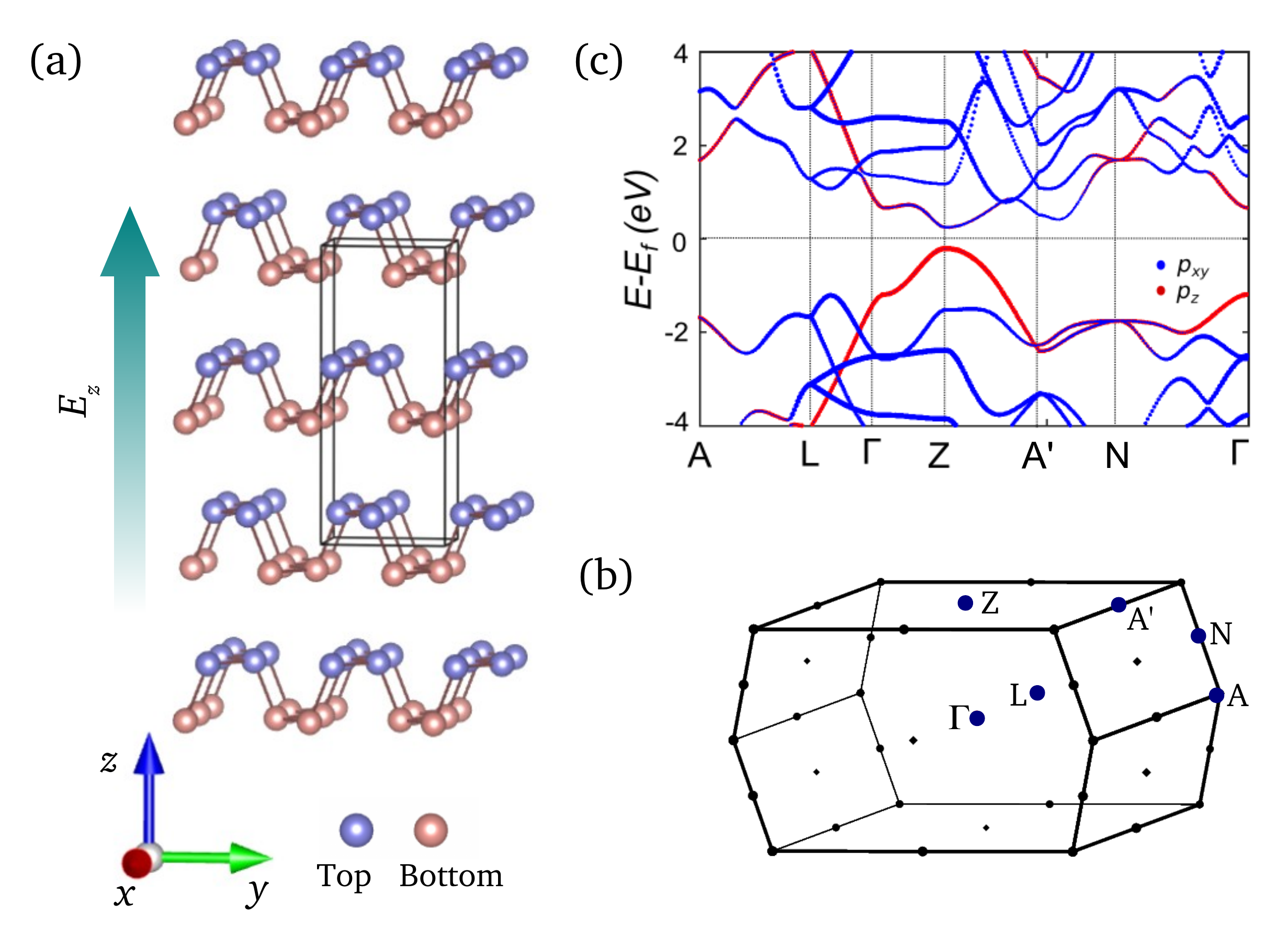} 
\caption{(a) Crystal structure of multilayer stack ($5$ layers) of phosphorene. Blue and brown balls represent the non-planner P layers and black box identifies the bulk conventional orthorhombic unit cell. A transverse electric field, $E_z$, is applied along the $z$ direction. (b) Brillouin zone for bulk black phosphorous with the  primitive orthorhombic unit cell  in which the high symmetry points are marked. (c) Bulk band structure of black phosphorous along the high symmetry directions in the primitive Brillouin zone. Sizes of blue and red markers denote the weight of projections onto $p_{xy}$ (predominantly $p_y$ near the CBM) and $p_z$ orbitals of P, respectively.} 	
\label{f1}
\end{figure}

Phosphorene has a puckered honeycomb structure with {non-symmorphic $D^7_{2h}(Pbmn,53)$ space group which is a subgroup of bulk black phosphorus orthorhombic $D^{18}_{2h}(Aema,64)$ space group}. The few layers ball-stick model of phosphorene is shown in Fig. \ref{f1}(a) where phosphorene layers are held together predominantly by weak van der Waals forces \cite{Hyd, doi:10.1021/acs.nanolett.5b03615}. Within a single phosphorene structure, each P atom forms three bonds with its neighbors and since P atoms have strong {\it{sp$^3$}} hybridization character, the three bonds have an arrangement similar to that of a tetrahedral configuration. This results in two atomic planes which are explicitly shown with blue and brown balls in Fig. \ref{f1}(a). The bonding between the atoms inside an atomic plane forms zigzag chain [along the $x$-axis in Fig. \ref{f1}(a)]  whereas it has an armchair structure between two atomic planes [along the $y$-axis in Fig. \ref{f1}(a)]. Due to such a lattice structure, phosphorene has the following symmetry elements: an inversion center $i$, a vertical mirror plane $\tilde {M}_x$ perpendicular to $x$, two-fold rotation axis along $x$-axis $C_{2x}$, two glide-mirror planes $\{\tilde{M}_y|\frac{1}{2}\frac{1}{2}\}$ and $\{\tilde{M}_z|\frac{1}{2}\frac{1}{2}\}$ and two screw axes $\{C_{2y}|\frac{1}{2}\frac{1}{2}\}$ and $\{C_{2z}|\frac{1}{2}\frac{1}{2}\}$ \cite{dresselhaus}. 

\section{Electronic structure of ultra thin films of black phosphorus}\label{QCE}

Before discussing the electronic properties of phosphorene thin films, we first calculate the electronic structure of bulk black phosphorous which consists of an infinite number of phosphorene layers.  As shown in Fig.~\ref{f1}(c), bulk black phosphorous is a direct band gap semiconductor with valance band maximum (VBM) and conduction band minimum (CBM) located at the $Z$-point of the bulk Brillouin zone. The valance and conduction bands disperse anisotropically around the $Z$-point with a nearly parabolic (linear) dispersion along $Z-A'$ ($Z-\Gamma$) direction. The calculated band gap with van der Waals correction is 0.44 eV. Though this bandgap value is slightly higher than the experimentally reported value of 0.31$-$0.35 eV,  it is in good agreement with the earlier DFT calculations \cite{P4,Dolui2015} incorporating van der Waals correction \footnote{The reported band structure is calculated using fully relaxed structural parameters, which are obtained using the PBE + vdw-DF2 energy functional. On calculating the relaxed structural parameters obtained using the PBE functional we obtain the lattice parameters to be $a=3.31$ \r{A},  $b=4.56$ \r{A} and $c=11.31$ \r{A}, and the corresponding bandgap is 0.29 (0.15) eV with (without) van der Waals correction -- consistent with Ref.~[\onlinecite{P4}]. Note that while there could be some uncertainty in estimating the lattice parameters and the band gap due to inherent limitations of DFT, the  insulating ground state of bulk back phosphorous  is correctly predicted.}. 
The local density of states analysis shows that VBM arises predominantly from the 3$p_z$ states, whereas CBM has a large contribution from 3$p_y$ and 3$p_z$ states of P.  

\begin{figure}[h!] 
\includegraphics[width=0.50\textwidth]{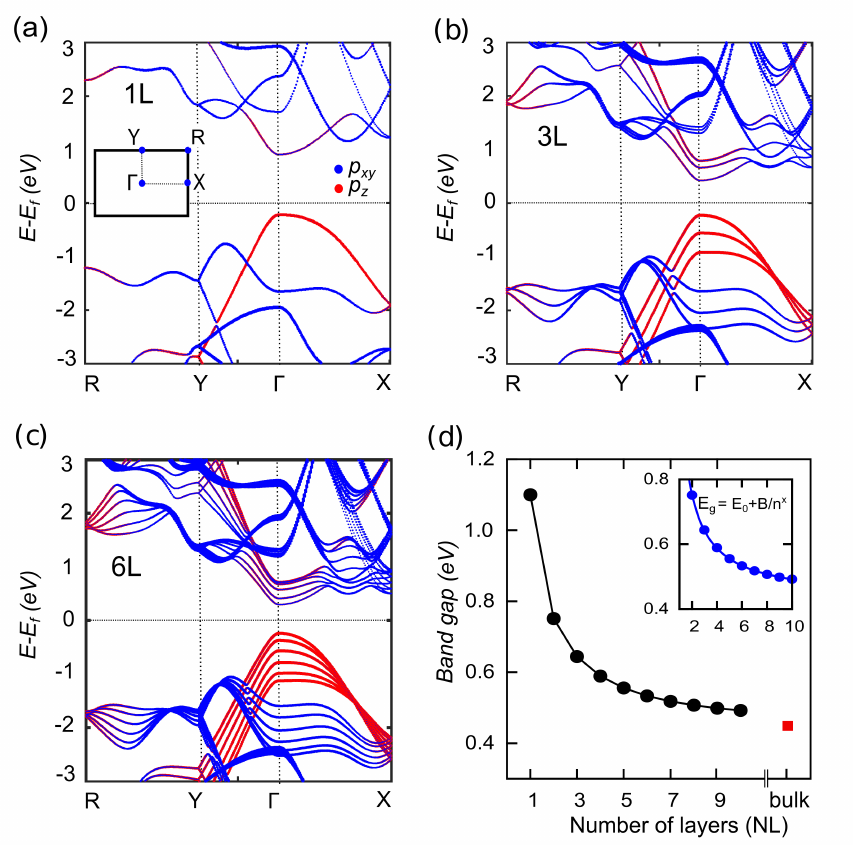}
\caption{Band structure of (a) 1L, (b) 3L, and (c) 6L thick films of phosphorene along the high-symmetry directions in the 2D Brillouin zone [see inset of panel (a)]. Sizes of various markers denote the weight of projections onto various atomic orbitals of P. (d) Band gap as a function of number of phosphorene layers (NL). The inset shows the band gap vs NL fitting curve (blue solid line) using the fitting function $E_g=E_0+B/n^x$ (see text for details).	
\label{f2}}
\end{figure}

Figure \ref{f2} shows the electronic structure of NL phosphorene. There is a dramatic increase in the band gap from bulk to 1L phosphorene which has much higher band gap of 1.1 eV [see Fig. \ref{f2}(a) and  \ref{f2}(d)]. This is a well-known phenomenon in low-dimensional systems and arises because of the quantum confinement of the charge carriers. Furthermore, we find that when the phosphorene monolayers are brought together to form a NL phosphorene structure, the degeneracy of the conduction and valence bands is lifted on account of the interlayer interaction between the phosphorene layers. This results in the appearance of more bands in the vicinity of the Fermi level. Notably, with the addition of 1L phosphorene into the NL structure, one extra band appears near the CBM and VBM (see Fig.~\ref{f2}) such that the total number of valance or conduction bands around VBM and CBM are equal to the number of phosphorene layers, NL. Since these bands repel each other, a monotonic decrease in the band gap occurs from 1L phosphorene to the bulk. There is a large decrease in the band gap from 1L to 3L (from 1.1 eV to 0.64 eV) beyond which the band gap decreases slowly and converges to the bulk value. Interestingly, the direct band gap nature at the $\Gamma$-point remains intact in NL phosphorene with increasing the thickness. This is unlike the case of MoS$_2$ where a direct to indirect band gap transition occurs with increasing the number of layers \cite{MoS2}.

To highlight the quantum confinement effects in NL phosphorene, we investigate the relationship between the band gap and the number of layers, NL, as shown in Fig.~\ref{f2}(d). A fitting to the layer dependent band gap ($E_g$) yields a power law fit of the form $E_g = E_0+B/n^x$, where $E_g$ is the band gap of NL phosphorene, $E_0$ is the band gap of bulk black phosphorous, and $n$ are the number of layers. The fitting parameters $B$ and $x$ are found to be 0.66 eV and 1.09, respectively. This clearly shows that band gap decreases slowly as compared to the  inverse square decay of the form $1/{n^2}$ as expected from the quantum confinement. 
The weak quantum confinement is a consequence of interlayer interactions (including van der Waals interaction) between the phosphorene layers \cite{doi:10.1021/acs.nanolett.5b03615}, which delocalizes the electrons in a given layer over the adjacent layers \cite{P3}. 

The energy dispersion in NL phosphorene in the vicinity of the $\Gamma$-point remains anisotropic with different band curvature along the $\Gamma-X$ and the $\Gamma-Y$ directions. This is similar to the case of bulk phosphorous bands and it reflects the anisotropy of underlying crystal structure. Note that the bulk Z-point folds back to $\Gamma$-point and thus the $\Gamma-X$ and $\Gamma-Y$ in NL phosphorene Brillouin zone lie along the zigzag and armchair directions, respectively, of the crystal structure [see Fig. \ref{f1}(a)]. In particular, the valance and conduction bands have a `Schr\"odinger' like parabolic dispersion along the $\Gamma-X$ (zigzag) direction whereas they have a `Dirac' like linear dispersion along the $\Gamma-Y$ (armchair) direction. Such an anisotropic band dispersion leads to anisotropic carrier effective masses. The estimated electron (hole) masses are  $m_e^*=1.2m_e$ ($m_h^*=8.64m_e$) and $m_e^*=0.17m_e$ ($m_h^*=0.16m_e$), respectively, along $\Gamma-X$ and $\Gamma-Y$ directions for 1L phosphorene. As we increase the number of phosphorene layers, the effective electron masses remain almost invariant along both $\Gamma-X$ and $\Gamma-Y$ directions whereas the effective hole masses show strong layer dependence along $\Gamma-X$ direction \cite{P4}. 

\section{Effect of a transverse electric field and the Dirac semimetal state}\label{EField} 
\begin{figure}
\includegraphics[width=0.5\textwidth]{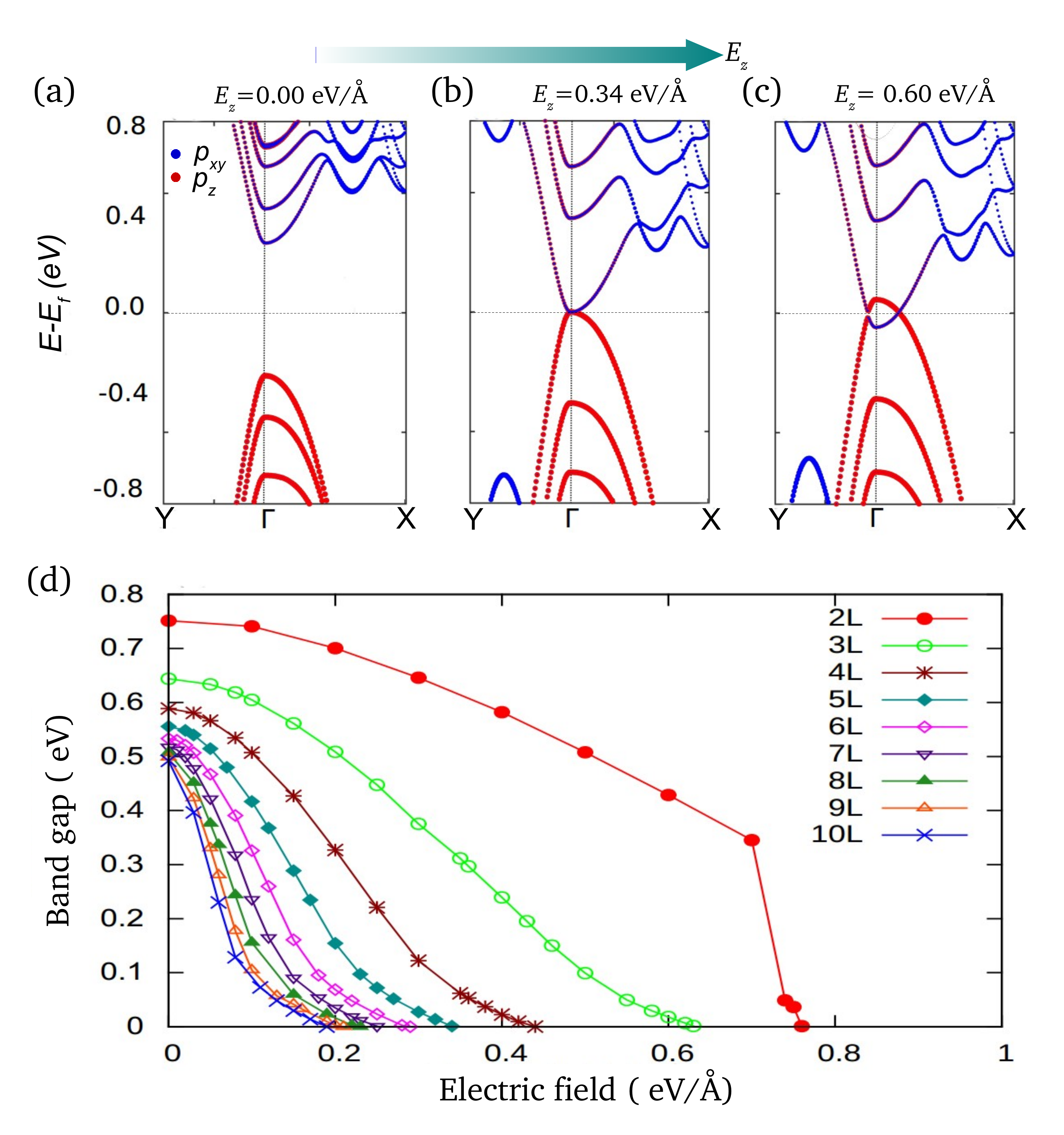} 
\caption{Electronic structure of 5L phosphorene film for various external electric field, $E_z$, values: (a) $E_z$ = 0.00 eV/\r{A}, (b) $E_z$ = 0.34 eV/\r{A}, and (c)  $E_z$ = 0.60 eV/\r{A}. There is a phase transition from normal insulator to Dirac semimetal state at $E_z$ = 0.34 eV/\r{A}, beyond which, an electric field tunable Dirac fermions state appears along $\Gamma-X$ line in 2D BZ. (d) Variation of band gap of phophorene thin films as a function of external electric field $E_z$. Various colors refer to different number of phosphorene layers (see legends). } 
\label{f3}
\end{figure}

We now discuss the effect of a transverse electric field $E_z$ on phosphorene multilayer structure and the resulting Dirac semimetal state. The existence of the potential difference between the two atomic layers in the puckered honeycomb structure of phosphorene naturally provides an advantage in controlling the band gap via an out-of-plane electric field. In addition the external electric field interacts strongly with localized atomic orbitals in phosphorene, changing the onsite energies as well as the hopping parameters which in turn have a significant impact on the resulting bandstructure \cite{Doh2016}.
Figure \ref{f3} shows the evolution of 5L phosphorene band structure with the strength of the external transverse electric field - $E_z$ (see Fig.~\ref{f1}). With an increase in $E_z$, the band gap decreases gradually while maintaining its  direct nature at the $\Gamma$-point and closes at a critical electric field - $E_c$, for which the VBM and CBM become degenerate at the $\Gamma$-point. The decrease in the band gap of multilayered phosphorene with $E_z$ arises mainly because of the stark effect. The applied electric field increases the potential difference across the layers and affects the real space distribution of valance and conduction states. Specifically, the real space distribution of valence and conduction states shifts in opposite direction to each other on account of the electrostatic interaction, leading to a reduced band gap. Note that although we have used 5L phosphorene to explicitly show the evolution of band structure with $E_z$ [in Figs.~\ref{f3}(a)-(c)], our main findings remain independent of the number of layers. Indeed we find that the band gap of NL phosphorene decreases with increase in $E_z$ and closes at a critical field, $E_z =E_c$. This is highlighted in Fig.~\ref{f3}(d) which explicitly shows the direct bandgap evolution for 2L$-$10L as a function of $E_z$. The band gap reduces rapidly with $E_z$ as we increase the number of phosphorene layers. Specifically, $E_c$ reduces from 0.75 eV/\r{A} for 2L to 0.19 eV/\r{A} for 10L. 
This is mainly because the potential drop across the layers increases rapidly with increasing the number of layers.

\subsection{Emergence of Dirac cones}
\begin{figure}[t!]
\includegraphics[width=0.50\textwidth]{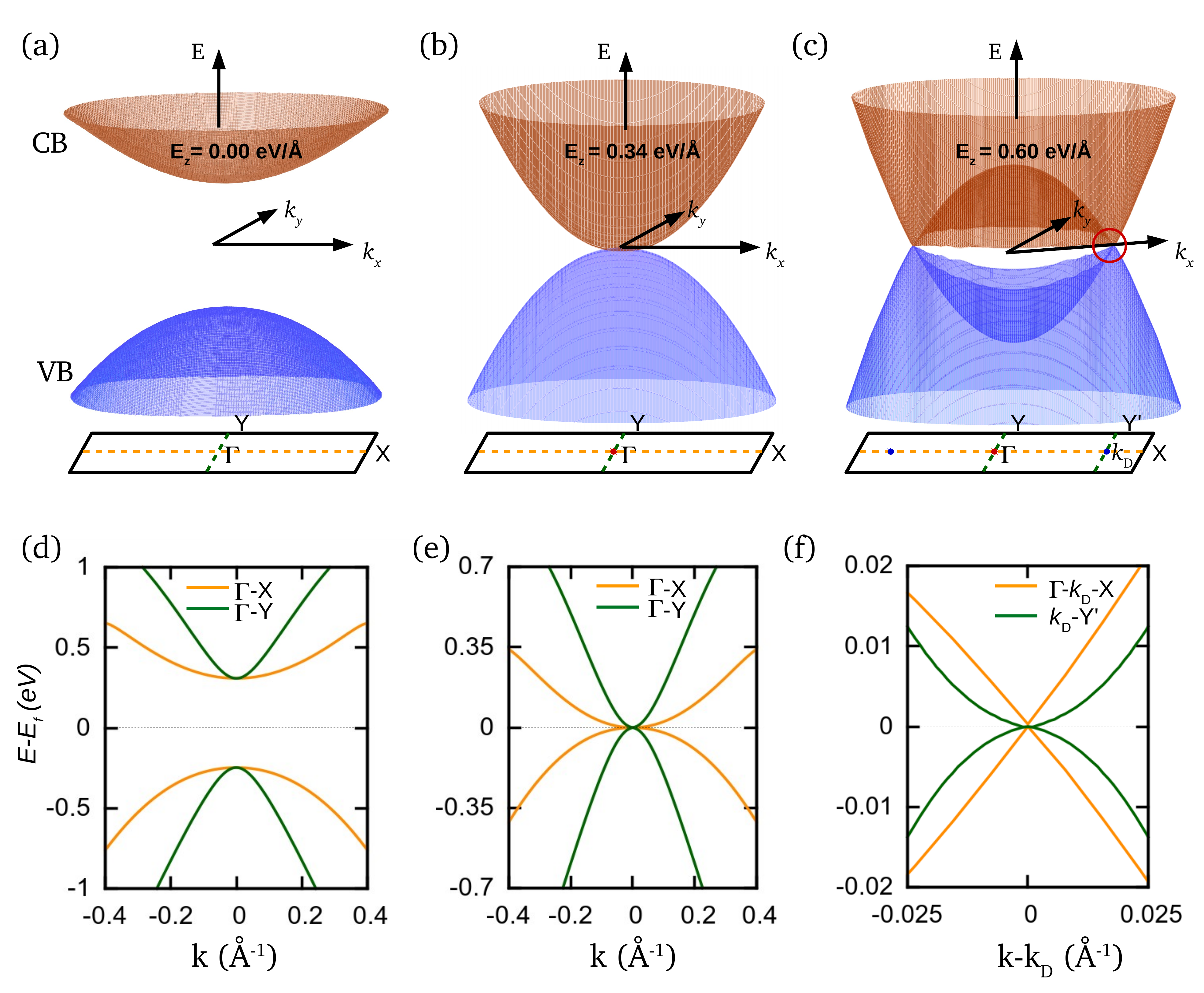} 
\caption{Evolution of valence band (VB) and conduction band (CB) of 5L phosphorene film with electric field in 2D BZ for (a) $E_z < E_c$, (b) $E_z = E_c$,  and (c) $E_z > E_c$. Panels (d) and (e) show energy dispersions of VB and CB along $\Gamma-X$ (orange color) and $\Gamma-Y$ (green color) directions for $E_z < E_c$ and $E_z = E_c$, respectively, near $\Gamma-$point. (f) Band dispersions along $\Gamma-k_{\rm D}-X$ (orange color) and $k_{\rm D}-Y'$ (green color) directions near the Dirac point ($k_{\rm D}$) for $E_z > E_c$ [see panel (c) for $k_{\rm D}$]. }
\label{f4}
\end{figure}

Interestingly, as we increase $E_z$ beyond $E_c$, a band gap starts opening at the $\Gamma$-point which increases monotonically with increasing $E_z$ [for example see Fig. \ref{f3}(c)]. The evolution of valence and conduction bands around the $\Gamma$-point with $E_z$ is reminiscent feature of band inversion observed in topological insulators. Indeed, by tracking the orbital character of VBM and CBM with the electric field, one confirms that the $p_z$ ($p_z + p_y$) states of phosphorous which constitute VBM (CBM) for $E_z <E_c$ invert their order and form CBM (VBM) for $E_z > E_c$ [see Fig. \ref{f3}(c)]. Furthermore,  we find  that although an inverted band gap opens at the $\Gamma$-point for $E_z > E_c$, the system remains gapless with a Dirac-type band crossing along the $\Gamma-X$ line at ${\bf k} = \pm {\bf k}_{\rm D}$, where we have defined ${\bf k}_{\rm D} \equiv (k_{\rm D},0)$ [see Fig.~\ref{f3}(c) and Fig.~\ref{f4}(c)]. The asymptotic band dispersion in the vicinity of this point is linear and anisotropic as shown in Fig.~\ref{f4}(f). A symmetry analysis suggests that this crossing point is indeed a Dirac point which is protected by the glide mirror plane $\{\tilde{M}_y|\frac{1}{2}\frac{1}{2}\}$. Note that a transverse electric field $E_z$ preserves this symmetry and since $\tilde{M}_y$ sends $y \rightarrow -y$, it leaves the $\Gamma-X$ line invariant under reflection. As shown explicitly in Ref.~[\onlinecite{kane2d}], any crossing point on this invariant line should switch the eigenstates of non-symmorphic symmetry $\{\tilde{M}_y|\frac{1}{2}\frac{1}{2}\}$ and therefore remain gapless.

Figure \ref{f4} illustrates the evolution of the valance and the conduction bands across the critical electric field near the $\Gamma$-point. For $E_z < E_c$, the valence and conduction band dispersion is linear along $\Gamma-Y$ direction and quadratic along $\Gamma-X$ direction. However, for $E_c> E_z$, there is a band inversion with the valence and conduction bands switching positions at the $\Gamma$-point, followed by the formation of a pair of Dirac points at $\pm {\bf k}_{\rm D}$.  The band dispersion around the Dirac points is linear along both the $\Gamma-{\bf k}_{\rm D}-X$ and the ${\bf k}_{\rm D}-Y'$ (line passing through ${\bf k}_{\rm D}$ and parallel to the $y$ axis) direction. Note that while the bands remain linear along $\Gamma-k_{\rm D}-X$ and $k_{\rm D}-Y'$ direction, an anisotropy in the carriers velocity is evident [see Fig. \ref{f4}(f)] because of the structural anisotropy of phosphorene. Furthermore, the Dirac points move continuously over the $\Gamma-X$ line towards zone boundary $X$ with increasing electric field strength.   

In the absence of SOC, the effective Hamiltonian in the vicinity of the Dirac point $\pm {\bf k}_{\rm D}$ can be expressed as 
\begin{equation} \label{eq1}
H \approx \left[ \hbar v_x (k_x \mp k_{\rm D}) \tau_x + \hbar v_y k_y \tau_y \right] \otimes s_0~,
\end{equation}
where $\tau_{x/y}$ denotes the $2\times 2$  Pauli spin matrices for the pseudo-spin degree of freedom, $s_0$ denotes the  $2 \times 2$ identity matrix in the actual spin space, $\otimes$ represents the direct product, and $v_x$ ($v_y$) denote the Dirac velocity in the $x$ ($y$) direction. Using the band dispersion resulting from Eq.~\eqref{eq1}, the inverted bandgap at the $\Gamma$-point can be approximated as $E_g = 2 \hbar v_x k_{\rm D}$.  
Based on our first-principles results, we find that $E_g \propto (E_z - E_c)$, $v_x \propto \sqrt{E_z-E_c}$, and  $k_{\rm D} \propto \sqrt{E_z-E_c}$ for $E_z > E_c$ [see Fig.~\ref{f6}(a) for $k_{\rm D}$]. These results clearly show that the Dirac-points as well as the associated carrier velocities can be modulated by an external electric field. 

\subsection{Impact of spin-orbit coupling and appearance of spin-polarized Dirac cones}
Since P is a lighter atom ($Z = 15$), it has weak spin-orbit interaction, which in turn has a very small effect on the electronic properties of NL phosphorene. However it plays a very significant role in presence of an external electric field $E_z$, leading to the emergence of four Dirac points.  The application of $E_z$ breaks the inversion symmetry of NL phosphorene as the two sides of NL phosphorene (along the external field) become inequivalent. As a result, the spin-degeneracy of states away from the time reversal invariant points, is lifted. In this way, the electric field provides a way to realize spin-polarized states in NL phosphorene.

\begin{figure}[t!]
\includegraphics[width=0.5\textwidth]{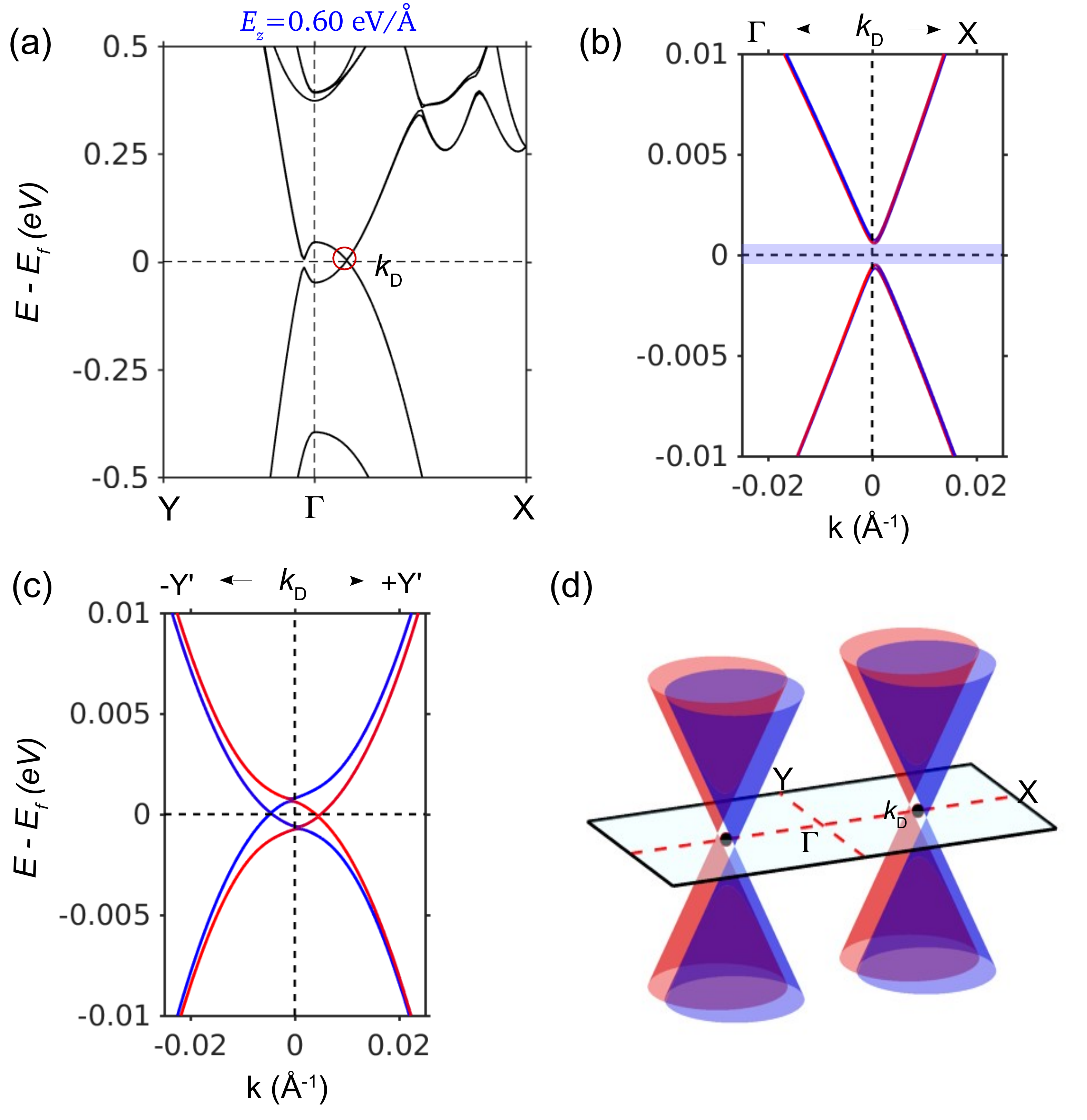} 
\caption{(a) Band structure including the SOC of 5L phosphorene film with $E_z$ = 0.60 eV/\r{A} along high symmetry directions in 2D Brillouin zone. Panels (b) and (c) show zoomed-in band structure along $\Gamma-k_{\rm D}-X$ and $k_{\rm D}-Y'$ directions, respectively. Each Dirac cone without SOC splits into two spin-polarized Dirac cones in presence of SOC with spins aligned almost along $+x$ (red color) and $-x$ (blue color) directions. Blue color stripe in (b) shows the SOC -induced band gap. Note that in panel (c), the central line corresponds to the mirror plane of the $\{M_y| \frac{1}{2} \frac{1}{2}\}$ symmetry, and  the reflection of the spins (which are almost aligned along the $x$ axis) about this mirror plane, is evident. (d) Schematic band dispersions of four spin-polarized Dirac cones in phosphorene with SOC for $E_z > E_c$. Red and blue colors represent spin orientated along $+x$ and $-x$, respectively.
\label{f5}}
\end{figure}

Figure \ref{f5} shows the band structure of 5L phosphorene with SOC for $E_z$ = 0.60 eV/\r{A}. It is evident from the zoomed-in band structure around the two Dirac points $\pm {\bf k}_{\rm D}$ [see Figs.~\ref{f5}(b) and \ref{f5}(c)], the spin-degeneracy of the bands is lifted and a band gap (of $\sim$ 3 meV), appears at  point ${\bf k}_{\rm D}$ along $\Gamma-k_{\rm D}-X$ directions. This is consistent with earlier findings where a SOC induced band gap is seen at the position of Dirac point without SOC\cite{nanolettntot,Dolui2015}. However, a fine $k$-mesh calculations near $\pm{\bf k}_{ D}$ reveals that although a SOC induced gap opens at $\pm{\bf k}_{\rm D}$, each of the previous Dirac points (without SOC) split into two, and the system still remains gapless along ${\bf k}_{\rm D}-Y'$ directions. Thus a total of four spin-polarized Dirac cones appear at  ${\bf k} = (\pm k_{\rm D},\pm k_{\rm SOC})$, where $2k_{\rm SOC}$ gives the splitting between the two Dirac cones at $k = (k_{\rm D},\pm k_{\rm SOC})$ [see Figs.~\ref{f5}(c) and \ref{f5}(d)]. A spin-texture analysis shows that the carrier spins for the two Dirac cones located at ${\bf k} = (\pm k_{\rm D},+k_{\rm SOC})$ are aligned almost along the $+x$ direction, while for the other two Dirac cones located at the ${\bf k} = (\pm k_{\rm D},-k_{\rm SOC})$, the spins are nearly aligned along the $-x$ direction. Specifically, the $x$ component of spin is found to be $0.95\sim 0.99 $ times $\hbar/2$ whereas $y$ component of spin is $0.00 \sim 0.20 $ times $\hbar/2$. A similar spin orientation of Dirac cones has been reported recently in 4L phosphorene under K doping \cite{Kdoped1}.  

\begin{figure}[t!]
\includegraphics[width=0.5\textwidth]{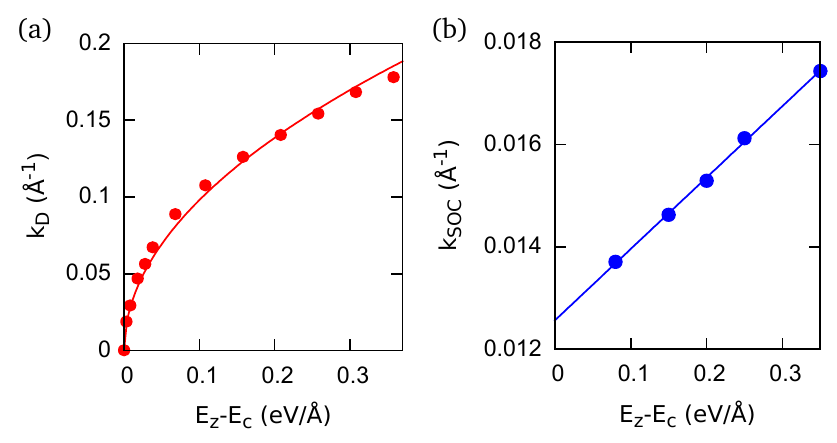} 
\caption{(a) $k_{\rm D}$ as a function of $E_z-E_c$. The red dots are based on DFT calculation, and the solid line is the fit $k_{\rm D} = \alpha \sqrt{E_z-E_c}$, where $\alpha = 0.31$ (eV\AA)$^{-1/2}$ based on GGA calculations.  (b) $k_{\rm SOC}$ as a function of $E_z-E_c$. The blue dots are based on DFT calculation, and the solid blue line is the fit $k_{\rm SOC} = k_0 + \beta (E_z-E_c)$, where $k_0 = 0.013$ \AA$^{-1}$ and $\beta = 0.014$ eV$^{-1}$  based on GGA calculations. Note that to accurately locate the Dirac point, we have used the increased SOC strength of $500$ percent. 
\label{f6}}
\end{figure} 

In principle the SOC can break the non-symmorphic symmetries and consequently the Dirac cones in a 2D lattice can be gapped out in presence of SOC. However, based on our first-principles results, an effective SOC Hamiltonian of phosphorene has the form 
\begin{equation} \label{eq2}
H_{\rm SOC}=\lambda \frac{\hbar}{2} \tau_y \otimes s_x~,
\end{equation} 
where $\lambda$ is the spin-orbit coupling constant \cite{Kdoped1}. 
It turns out that this $H_{\rm SOC}$ term commutes with the glide-mirror operator $\{\tilde{M_y}|\frac{1}{2}\frac{1}{2}\} = i\tau_x \otimes s_y$ and thus does not break glide-mirror symmetry. This can be further verified from the first-principles results, as shown in Figs. \ref{f5}(b) and \ref{f5}(c), where the spin degeneracy remains almost intact along $\Gamma-X$ direction due to the mirror-symmetry. The spin-degeneracy is however lifted away from this line (for example along $\Gamma-Y'$ direction) because of SOC and therefore, the Dirac points shift their position along the ${\bf k}_{\rm D}-Y'$ direction to lie on to generic $k$-points. Since the non-symmorphic symmetry $\{\tilde{M_y}|\frac{1}{2}\frac{1}{2}\}$ which is respected by the form of the the SOC in Eq.~\eqref{eq2}, forbids the presence of mass term, the Dirac cones in presence of SOC are still symmetry protected \cite{kane2d,Kdoped1,Doh2016}.
Based on Eqs.~\eqref{eq1}-\eqref{eq2}, it can be easily shown that $k_{\rm SOC} = \lambda/(2 v_y)$. While $\lambda$ is a constant, we find that $v_y \propto (E_z-E_c)^{-1}$, and as a  consequence $k_{\rm SOC} \propto (E_z-E_c)$ [see Fig.~\ref{f6}(b)]. Thus we find that the  location as well as the Fermi velocities of the four Dirac cones can be tuned by changing the strength of the external electric field.

Based on the analysis above, we emphasize that NL phosphorene remains in its semimetal state in presence of SOC and an external electric field, in contrast to earlier studies \cite{nanolettntot,Dolui2015}. Furthermore, the spin resolved Dirac cones in phosphorene in presence of SOC and a transverse electric field, are located at the generic $k$ points, showing the realization of the unpinned type-II Dirac cones proposed recently by Lu {\it et al.} [\onlinecite{Lu2016}]. 

We have discussed that in 5L phosphorene a critical electric field of $\sim 0.34$ eV/\AA~ is sufficient to induce the phase transition. This critical electric field could be large to achieve in normal experimental conditions and thus, it important to discuss the ways of reducing the critical electric field. As shown in Fig. \ref{f3}(d), the critical electric field decreases with increasing the number of phosphorene layers. Experimentally, the high quality phosphorene sample lies in the range of of 5-10 nm, which corresponds to 10-20 phosphorene layers\cite{P1,P0,nanolettntot}. Therefore one could use 10-20 phosphorene layers to realize phase transition at smaller critical electric fields. An alternative way to emulate the large transverse electric field is via giant Stark effect using the chemical doping\cite{Kdoped1}. This method has been recently demonstrated experimentally as an effective way to tune band gap and drive few layers phosphorene into a semi-metallic state with Dirac fermions\cite{Kim723}.

\section{Summary and conclusions}\label{concl}
We have investigated the electronic properties of NL phosphorene within the {\it ab-initio} density-functional theory framework. Phosphorene is found to be a versatile material in which the electronic properties can be tuned by varying the thickness as well as by applying a transverse electric field. More importantly, we have shown that increasing the electric field beyond its critical value induces a band inversion at the Brillouin zone center. However, the inverted valence and conduction bands cross at only two points in the Brillouin zone along $\Gamma-X$ directions forming a pair of Dirac cones, which are protected by the glide mirror plane symmetry of the phosphorene lattice. The SOC splits these two Dirac cones into four fully spin-polarized Dirac cones which lie on  generic  $k$-points on the 2D Brillouin zone. The new Dirac cones have unique spin-textures where spins are aligned almost along $x$-directions. We find that the position of the four Dirac cones, as well as the associated carrier velocities, are electrically tunable. Our results establish that NL phosphorene realizes an electric field induced phase transition from normal insulator to the Dirac semimetal state.

\bibliography{BP_DS_final}

\end{document}